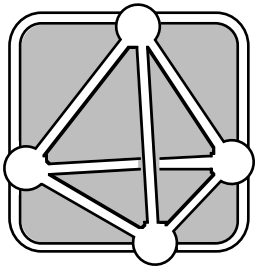

# ASK•IT / A2L: Assessing Student Knowledge with Instructional Technology

**Robert J. Dufresne, William J. Gerace, Jose P. Mestre, and William J. Leonard**
*Department of Physics, University of Massachusetts, Amherst, MA 01003–4525*
contact: dufresne@physics.umass.edu

There is a persistent mismatch between the goals of physics instruction and everyday classroom practices. Physics teachers at all levels typically have laudable, long-term goals for their students; they want to help students develop a deep understanding of concepts, a willingness to apply principles to problems and situations, and an appreciation of physics as a process of inquiry. Yet their classroom practices often drive students toward undesirable, short-term goals; students become overly focused on "right" answers and fail to learn the ideas, skills, and principles needed for conceptual understanding and flexible problem solving.

One reason for the mismatch between goals and outcomes is the way many teachers assess their students. Periodic, cumulative exams have a tendency to concentrate on information and low level skills. These exams can be high stakes and stressful, and both teachers and students can become overly focused on spending all available mental resources on preparing for them. And perhaps the worst part of the cycle is that the information gleaned from the exams about what students do and don't understand is not used to modify learning modes or instructional practice.

Teaching is communication, and we believe that teachers should go into the classroom as much to learn as to teach. One way to break out of the cycle of preparing for and giving cumulative exams is to use *formative* assessment throughout the instructional process. Formative assessment informs. It informs the teacher about what students think and about how they think. It informs students about what and how their classmates think. It informs students about how they themselves think. It improves communication and learning in the classroom.

Formative assessment is low stakes and low stress. It is done often. The focus is not on the "right" answer, but on the distribution of answers and the reasoning behind each one. It shows students that even if they are in the minority there are others who think similarly. It helps teachers tailor instruction to fit the needs of their students.

However, formative assessment can be time-consuming and it can be difficult and awkward to collect and organize the answers from the whole class. Thus, instructional technology becomes a useful component for ensuring that everyone participates, for collecting, sorting, and displaying answers efficiently, and for easing the transitions between different activities during a lesson.

The ASK•IT / Assessing-to-Learn (A2L) project is an attempt to bring a strategic approach to learning, instruction, and communication. ASK•IT / A2L seeks to integrate formative assessment and instruction in the physics classroom at both the high school and college levels. The project and the approach are the result of years of experience in teaching with and without instructional technology, as well as years of experience in cognitive research. Its goals are:

- to facilitate the use of formative assessment in the classroom;
- to show how technology can make the classroom more interactive and to show how it can support teachers rather than replace them;
- to help teachers assess students' cognitive development, rather than assessing their own delivery of information or students' acquisition of knowledge;
- to study the role of technology in the classroom and the choice of assessments on student learning, reasoning, problem solving, and communication.

While an interactive classroom is more engaging and more fun, it can also be difficult for students to learn how to learn in an interactive environment. They need to develop new skills and coping strategies, such as listening, explaining, observing, interpreting, and summarizing. In the part I of this document, we present a structure to help discuss the new mindsets for students in an ASK•IT / A2L classroom. We have broken down the desirable tendencies of students into six basic and six advanced "habits of mind". By focusing on these good habits of mind, we hope to encourage students to actually <u>seek</u> to develop the skills and knowledge they need for deep conceptual understanding, enhanced critical thinking, and proficient problem solving.

Teachers also have new roles to learn. In part II, we present a set of cognitive goals for students. These goals are cognitive stages that most students need to follow in order to develop desirable knowledge and skills. By shifting the focus to these five stages of development, teachers can avoid the tendency to become overly concentrated on short-term goals. Teachers can therefore concentrate instead on solving the "problem" of instruction by focusing on student processes rather than the "rightness" of students' answers.

Rather than a lesson or an activity, the smallest "unit" in the ASK•IT / A2L project is an *item*, which is a question, problem, or task given to students to work on individually or in groups. Predictably, the classroom dynamic is very different when the core is an item rather than a lecture or a lab activity. To explore this realm, we show in part III how a typical lesson might be organized around an item.

Often it is necessary for a teacher to create an item on their own. In part IV, we present our *model-based design paradigm*, which we use for all of our development projects. We also describe some tips for avoiding common pitfalls in item creation and ways to match up cognitive goals with habits of mind, which we use as instructional modes for item creation.

The habits of mind are useful not only for helping students learn physics; they are also useful for helping both students and teachers pay more attention to thinking about, discussing, and learning about learning, thinking, and communication. In part V, we apply the habits of mind to the *metacognition* of teachers and students.

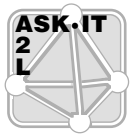

# PART I.
# New Attitudes for Students

The ASK•IT / A2L project encourages students to have a different mental attitude toward learning and instruction. To capture the essential elements of a positive mental attitude, we have identified 12 good "habits of mind" that students should strive to develop. These will help students not only cope with the steady stream of questions, interactions, and discussions characteristic of the ASK•IT / A2L classroom, but also help them to develop life-long skills.

## Habits of Mind

A habit of mind is a natural tendency or willingness as applied to mental processes. We know that students have certain "bad" habits of mind that can lead to failure. We would like to encourage "good" habits, and we believe that eventually these good habits of mind, when applied to activities and experiences, will lead to desirable learning outcomes, such as the development of critical-thinking skills.

Both students and teachers can have "bad" habits of mind, which can lead to limited growth: Students' tendencies can lead to unhealthy reliance on the memorization of facts and formulas, reduced engagement, and a lack of self-awareness. Teachers' tendencies can lead to unnecessary reliance on lecturing, avoidance of meaningful interaction with students, and a reduced ability to modify their approaches. Both can be overly attached to finding the "right" answer.

Encouraging "good" habits of mind in both students and teachers can lead to intellectual growth in both. Students can grow not only in their knowledge of physics concepts and problem solving, but also in their appreciation of themselves, their strengths and weaknesses, and learning in general. Teachers can grow not only in their knowledge of instruction and pedagogy, but also in their appreciation of teaching and the skills needed to teach. In other words, students are trying to develop the ideas and skills they need to understand and solve physics problems, while the teacher is trying to develop the ideas and skills they need to understand and solve the problem of instruction. And students are becoming more aware of themselves as learners and students, while teachers are becoming more aware of themselves as learners and teachers.

**"Basic" Habits of Mind.** We believe that certain habits of mind correspond to particular mental processes needed for active engagement and intellectual growth. The first set of these are referred to as "basic" because they are slightly more fundamental than those labeled as "advanced". They are also more widely applicable, because they can be encouraged during all stages of learning. The basic habits of mind are listed to the right. They will be explained in greater detail later.

> **BASIC HABITS OF MIND**
> Seek alternative representations
> Compare and contrast
> Explain, describe, draw, etc.
> Predict & Observe
> Extend the context
> Monitor and refine communication

> **ADVANCED HABITS OF MIND**
> Generate multiple solutions
> Categorize and classify
> Discuss, summarize, model, etc.
> Plan, justify, and strategize
> Reflect, evaluate, etc.
> Meta-communicate

**"Advanced" Habits of Mind.** The "advanced" habits of mind are more easily applied to later stages of learning. However, this does not mean that they are less valuable. In fact, to reach the stage of being a proficient, concept-based problem solver, students must have begun to develop all of these habits of mind. They are listed to the left. These also will be explained in greater detail in the next section.

## Applying Habits of Mind to the Classroom

Beginning students come into our classrooms with a wide variety of ideas about physics, about themselves, about instruction, about learning, about everything. Many of these "preconceptions" interfere with learning and inhibit each student's ability to develop desirable critical-thinking and problem-solving skills. Unless students confront any mismatch between physics ideas and their own conceptions, they will never understand the concepts and principles on which physics is based and they will never be able to solve physics problems outside of a limited context. Unless students develop analysis and reasoning skills, they will continue to use superficial approaches to solving physics problems. Their problem solving will remain algorithmic and reach barely beyond the level of equation and symbol manipulation.

We have a different vision of student learning and problem solving. We see students developing the skills needed to solve a wide range of problems efficiently and successfully. We see them structuring their knowledge for rapid recall and flexible application to new situations. We see students improving <u>both</u> their problem-solving and their conceptual understanding.

Ironically, these goals are achieved without focusing on either problem-solving tasks or rote memorization. Instead, we use a variety of educational experiences that stimulate beneficial mental processes. Learning occurs because the students are actively engaged in making sense of these experiences, communicating their ideas to classmates and to the teacher, and monitoring their own progress.

Each of the habits of mind has a role in prompting students to stay actively engaged. By encouraging each of these habits, we simultaneously help students learn concepts, develop skills, structure knowledge, and improve their own habits of mind.

**Seek alternative representations.** A representation can be algebraic, graphical, pictorial, physical, or verbal, just to name a few types. When students have command of many different representations, they are more likely to use them to communicate ideas (e.g., with classmates) and they are better able to analyze physical situations before solving problems. Having students interpret, translate between, and use a variety of representations helps them learn new ideas, distinguish new ideas from old ones, and relate ideas to each other.

**Compare and contrast.**  Good students are always seeking patterns.  Other students must be prompted to look for patterns.  When students look for and perceive similarities and differences between situations, they are more likely to distinguish concepts and appreciate the role of principles in analysis and problem solving.  They are also better able to interpret limiting cases and analogical reasoning to check their answers after problem solving.  However, to avoid oversimplified generalizations, teachers should be careful to avoid situations with unintended similarities.  For instance, students often think that the normal force always points vertically upward, because this was true in all of the examples used to introduce the concept.

**Explain, describe, draw, etc.**  Good teaching is often good communication—both among students and between students and teachers.  Good communication means that everyone is more aware of what everyone else is thinking, and for this students must be willing to describe their observations and experiences, explain their reasons behind any answers they give, and in general be open to letting others see the inner workings of their minds.  Students who have the confidence to display this habit of mind do not necessarily know all the answers or have the best explanations, but they are often the quickest to learn, because they are willing to commit to an answer or a line of reasoning and learn from their mistakes.

**Predict & Observe.**  Once students are more willing to communicate and are becoming more aware of their models of the physical world, they are ready to apply their models explicitly to new situations, defend their predictions, and compare their predictions with observations.  Many students are not comfortable with this process and would rather wait to see what the result of the observation will be.  What they don't realize is that they cannot as easily confront the flaws in their models, become self-aware, or learn unless they are willing to commit to an answer, explain it, and reflect on it.

**Extend the context.**  Each of us has a "safety zone" of thought, interaction, and experience, and it can be a little frightening to go outside of this zone.  But learning occurs necessarily at the boundary of this safety zone, so students benefit from being willing to consider and analyze unfamiliar situations.  When students are willing to extend the context, they are more likely to consider special and limiting cases as well as analogies.  Having students explore a range of contexts helps them to distinguish relevant features from irrelevant ones, and also helps them to avoid oversimplified generalizations.

**Monitor and refine communication.**  The last of the "basic" habits of mind is to pay attention to communication and to take steps to improve it when necessary.  In other words, students need to take greater responsibility in the communication process, whether it is between them and their classmates or between them and their teacher.  Because one of the premises of ASK•IT / A2L is that instruction should be a bi-directional communication process, this habit of mind has consequences for all the other habits of mind.

**Generate multiple solutions.**  Students often have at most one way of solving a problem or answering a question.  And if they are stuck, they usually don't know how to get themselves unstuck. Even simple problems and questions usually have many valid solutions and explanations, yet students perceive that there is only one "right" way and one "right" answer.

Successful problem solving often results from making choices about which method will be most easily applied to the problem situation. Having more than one way of solving a problem or answering a question gives students the opportunity to check their answer themselves without needing to refer to an authority (e.g., the textbook or the teacher) for confirmation. Knowing that there are many ways of solving problems and answering questions means that students with different skills and learning styles can succeed where before only students with good algebraic skills could succeed.

Having students solve the same problem in a new way sensitizes them to different approaches and helps them learn new principles and prioritize approaches. For instance, students who have learned how to solve problems using dynamics are sometimes reluctant to use energy ideas—even if energy seems easier to the teacher—simply because they are more comfortable with dynamics, not because they have made a reasoned decision. Therefore, let students solve a problem using dynamics in order to get the answer, then have them solve the same problem using energy (or momentum, etc.), and finally have them compare the results.

**Categorize and classify.** Looking for patterns, breaking up ideas, situations, or problems into categories, and naming them are useful tendencies, because they can lead to understanding, structured knowledge, and improved memory recall. By having students categorize and classify, they learn the value of concepts and principles for organizing knowledge. By exposing students to different classification systems, they learn to distinguish ideas.

A useful activity for students is to give them members of a group and have them find the feature that unifies them or that is common among them. For instance, the list might include *acceleration*, *velocity*, *force*, and *momentum*. The unifying concept is that they are all vectors. A variation is to then give students something that is <u>not</u> in the group, such as *speed*.

**Discuss, summarize, model, etc.** Higher order thinking skills are developed when students are willing to discuss, listen, seek clarification, ask questions, summarize, paraphrase, distill, and model. In short, students must interact in order to learn. Many students, however, are inclined to "tune out" what their classmates are saying and wait for the teacher to distill the information down to its essential content. Students often don't recognize when a classmate has given a valid explanation and they cannot determine what's wrong with an invalid one. These students are wasting valuable learning experiences, because they are sacrificing high level skills in favor of low level skills.

A discussion is most useful when the teacher does as little as possible to sustain it. The teacher provides the common experience around which the discussion revolves, but students should be the primary participants. They should be encouraged to seek clarification when they don't understand something and to debate ideas when they disagree with someone. Students should be insisting on agreed-upon definitions and self-consistency, and they should be considering alternative interpretations rather than jumping to conclusions. They should be learning how to be facilitators and moderators, rather than always assuming that the teacher will be responsible for these roles.

**Plan, justify, and strategize.** A *strategy* is defined to be a principle chosen to be applied to a problem, a justification of its validity and appropriateness for the particular problem situation and question, and a plan for applying the chosen principle and getting an answer. Forward-looking approaches are highly valued by scientists and teachers of science, but they are seldom encouraged in students.

Thoughtful engagement while problem solving is nearly synonymous with planning, justifying, and executing strategies. Students with this habit of mind think about where they are going, and they want to be able to recognize when they have arrived. By having students provide plans, justifications, and/or strategies for problems, teachers can encourage this habit of mind as well as promote conceptual understanding, knowledge structuring, and development of analysis, reasoning, and problem-solving skills.

**Reflect, evaluate, etc.** An essential habit of mind is reflection, along with all of its related higher order thought processes, such as evaluating, integrating ideas, generalizing, etc. Most students have not developed this habit of mind because they have not been asked to do so and they have not been given time to do so. Adults generally can reflect on activities and experiences while they are doing them; adolescents generally cannot. For instance, while reading for content, an adult will typically check for meaning, consistency, and interrelatedness. If they don't know a term, they will interrupt what they are doing, look up the term in a dictionary, and return to their reading. Students are less able to do this. They must be given time to explicitly reflect on their experiences, as well as time to verbalize their reflections. Because students do not usually appreciate the value of reflective thought, the activity must be made an integral part of instruction.

**Meta-communicate.** Perhaps the most desirable and highest level habit of mind is a desire and willingness to communicate about learning, which we refer to as *meta-communication*. Without this habit of mind, students become short-sighted and overly focused on short-term goals, such as getting the right answers to every question. To meta-communicate is to talk about learning styles and outcomes, about concepts and principles and their roles in problem solving, about structuring knowledge and its purpose, about the usefulness of learning physics or math or any other subject, about how different individuals have different learning speeds and approaches.

Meta-communication is also a mode of instruction that teachers can use to monitor student motivation and the classroom dynamic. By stepping outside the day-to-day activities, teachers can address critical issues and as needed modify accepted routines and modes of interaction. It's like a giant "re-set" button that teachers can use to get the class back on track.

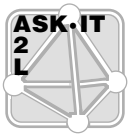

# PART II.
# New Roles for Students

In the ASK•IT / A2L classroom, students do not sit passively, drifting into and out of attention, taking notes that they can pore over at a later time. Instead, they have specific tasks or *items* that encourage them to apply and develop good habits of mind.

## Cognitive Goals

Rather than focusing on extremely short-term goals (e.g., Did everyone finish the homework last night?) or long-term goals (e.g., Who can flexibly apply physics concepts to a broad range of contexts? Who can pass the AP test?), we recommend focusing on intermediate goals. In particular, instead of focusing on problem-solving goals, we recommend substituting cognitive goals. These goals break down the development of knowledge and skills into five stages, all of which are needed for expert-like problem solving. They are listed to the right. Note that only one of the five stages refers specifically to problem solving. The rest are precursors to efficient, expert-like learning and problem solving.

**STAGES OF COGNITIVE DEVELOPMENT**
Explore, define, and hone concepts
Link and cluster concepts, operations, and procedures
Develop analysis and reasoning skills
Develop concept-based problem-solving skills
Organize, prioritize, and structure knowledge

## Applying Cognitive Goals to the Classroom

These goals can be applied to individual topics—such as *kinematics*, *interactions*, *Newton's laws*, and *conservation of momentum*—as well as to subtopics—such as *force*, *acceleration*, or *kinetic energy*. By breaking down the cognitive growth of students into these stages, it is easier to design educational experiences for them and to monitor their progress.

**Exploring, defining, and honing concepts.** Students come into the classroom with many deeply held ideas about the physical world, many of which survive instruction, and often co-existing with more formal, scientific ideas. To make any impact on students' learning, students must confront their prior conceptions and see how they match up, not only with physics concepts, but also with their own experiences.

The first stage of cognitive development is to have students become aware of their prior conceptions, to have students learn the scientific definitions, and then to have students begin to refine the definitions through a variety of activities and experiences. Like all of these stages, students do <u>not</u> need to fully understand concepts before they are ready to move on to the next stage of development. The later stages are essential for deep conceptual understanding, so spending too much time in this stage will actually inhibit students' intellectual growth.

**Linking and clustering concepts, operations, and procedures.** The same is true for the pacing of topics and subtopics. For instance, it is not necessary to finish *velocity* before moving on to *acceleration*, and students will deepen their understanding of both ideas by reflecting on how they are related. This is what we mean by *linking* concepts. It is also useful to cluster concepts together using physical laws and principles, such as Newton's laws and conservation of energy.

Operations and procedures are basic actions that students need to be able to perform to analyze, reason, and problem solve. For instance, they need to be able to compute the kinetic energy, draw a free-body diagram, solve for a particular variable in an equation, find the components of a vector, and distinguish between internal and external forces on a system. These actions, too, need to be linked to each other and to concepts so that they are fully accessible to students.

Most students will not link and cluster on their own, so teachers need to encourage students and monitor their progress through this stage. As mentioned before, students do not need to complete this stage before moving on to the next, and the later stages will actually increase students' appreciation of the value and importance of the interrelation of concepts, operations, and procedures.

**Developing analysis and reasoning skills.** Students must grow to appreciate the role of conceptual analysis and reasoning in problem solving. To them, solving a problem involves simply finding the right equation and then identifying and isolating the desired unknown. Students in this problem-solving mode are often not manipulating ideas, but manipulating symbols. Difficult problems are those for which two or more equations are needed. Students do not usually keep track of the assumptions and conditions associated with many derived relations, such as kinematic equations for constant acceleration. There is usually little or no consideration of whether the equations they use are valid or appropriate for the given situation.

Before students can solve problems using concepts, they must learn how to analyze and reason using concepts. Students can start by learning how to apply principles and laws to physical situations. Students can compare quantities rather than compute them. They can discuss the features of a situation most relevant for understanding it. They can listen to their classmates' explanations, evaluate them, construct arguments and counter-arguments, ask questions, and debate points of view. Students can also look for the representation that makes the analysis or reasoning simplest. They can look at one situation from many different perspectives, learning as much as they can from it before moving on to another situation.

**Developing concept-based problem-solving skills.** While students are developing a deep understanding of concepts, an ability to perform needed operations and procedures, and an appreciation of the value and importance of conceptual analysis, they can also start to create *strategic* knowledge elements, which are used to coordinate the problem-solving process. For instance, students need to know <u>when</u> to apply the Impulse–Momentum Theorem. Strategic knowledge helps students sort ideas and features into those that can be ignored and those that cannot. It guides students toward those operations and procedures needed to solve the given problem.

Students need practice developing strategic knowledge, but this does not mean they should do lots of problems. The essence of problem solving is critical thinking—the application of habits of mind to a problem or situation. Students, therefore, need to reflect on problem solving and to categorize problems. They can search for alternative solutions to problems they have already solved and compare solutions. They can attempt to <u>plan</u> their approaches to solving a given problem—without solving it first!—and then justify their proposed plans using concepts. These are difficult tasks for students, but they can be valuable experiences for extending their problem-solving abilities.

One important feature of this stage is for teachers to choose problems—to analyze, categorize, compare, or strategize—that require a conceptual approach. Giving too many problems that can be solved using equation manipulation will only stimulate undesirable student behavior and promote superficial learning. The problems do not need to be difficult or complicated. The best problems are actually those that are easily solved with a concept-based approach.

While students are developing their problem-solving skills, keep in mind that they should be improving in the other stages as well. Students' conceptual understanding should be deepened by problem solving. The number of interconnections between concepts, operations, and procedures should continue to grow. And their appreciation of the role of conceptual analysis and reasoning skills should increase as a direct result of successful problem solving.

**Organizing, prioritizing, and structuring knowledge.** Although the ultimate goal and measure of success for most physics courses is problem-solving proficiency, to us, there is still one stage left. The problem-solving skills are developed within a relatively limited context, and without structured knowledge, students will not have deep conceptual understanding and they will not be able to flexibly apply concepts and principles to unfamiliar situations and real-world problems.

The first step of this stage involves organizing ideas into different classes, such as concepts, physical laws and principles, definitions, empirical laws, derived relations, operations and procedures, mathematical principles, and problem-solving techniques. Then these ideas can be prioritized according to how widely useful and applicable they are. For instance, physical laws and principles are typically more widely useful than concepts and empirical laws. The last step is interconnecting the ideas into a hierarchical structure, with the most widely applicable ideas being used as umbrella concepts for less important ones.

During this stage, students are reflecting on and evaluating their experiences. They are integrating ideas and contexts that are seemingly unrelated. They are forming generalizations and models. They are talking about learning and about knowledge, about how they learn and about how they think. They are self-directed, self-aware individuals.

# PART III.
# New Roles for Teachers

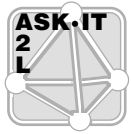

Teachers, too, must adapt to a new classroom environment. They must learn how to run an ASK•IT / A2L classroom: choosing items that will keep students engaged, making sure that students have the right amount of time to work on items, easing the transitions from question posing to individual or group work to class-wide discussion and wrap-up, and monitoring student progress. Most teachers will need to develop some new skills, as well as new habits of mind.

The ASK•IT / A2L classroom has a different emphasis than other classrooms. The focus is shifted to communication and modeling—students communicating with each other and with the teacher, students modeling themselves, each other, and the physical world, and the teacher modeling themselves and the students. Shifting the focus is not always easy, because both students and teachers often have habits of mind that inhibit any meaningful change from occurring. Three keys by which the focus can be successfully shifted are: (1) to ask students questions or to give them tasks; (2) to have students work in collaborative groups as much as possible; and (3) to use technology to facilitate real-time feedback.

## Formative Assessment

*Formative* assessment is assessment that informs—it informs teachers about what students think; it informs students what their classmates think; it informs individuals what they themselves think. It promotes awareness and communication in the classroom. The emphasis is on finding out what isn't known, and it tells teachers and students early enough in the course that something can be done about it. If done well, it is extremely low-stakes. If done often, both students and teachers can modify their approaches based on the outcome. In this way, formative assessment can be a valuable learning experience for students and teachers.

*Summative* assessment is different. It is used to evaluate—in the sense of assigning a value label. It is high-stakes. The emphasis is on finding out what is known. Both students and teachers tend to focus on the score rather than the details. It is done at the end of a topic, and thus, it cannot generally be used as a learning experience.

**The question cycle.** One way in which formative assessment can be used is called the *question cycle*. The typical class period can be broken down into two or three cycles of six phases, as listed to the right. Although this cycle does not need to be followed strictly at all times, it is a good starting point, especially for teachers who are new to conducting an interactive classroom. The cycle makes relatively efficient use of class time by always asking a question first. Discussion and wrap-up can then be shortened or lengthened to take into account students' answers.

| QUESTION CYCLE |
| :---: |
| Present an item |
| Let students work individually or in groups |
| Collect answers |
| Show histogram |
| Discuss as a class |
| Wrap-up |

*Presenting an item*

Although the act of "presenting" an item is straightforward, the relevance of this phase is that it is first. There is no lecturing needed beforehand; students merely answer the questions based on their current knowledge and understanding, and preferably, without worrying about whether their answers are right or wrong. However, the choice of question, problem, or task is critical. There should be a variety of types of items, and they should vary in their subtlety, abstractness, and level.

It is often a good idea to let students read and interpret the item themselves, without any immediate intervention from teachers. All questions contain some level of ambiguity, and it is often the case that the ambiguity leads to a range of defensible responses. By leaving the ambiguity there, teachers can help discourage students from focusing on the "right" answer and can help them develop interpretational and communication skills.

*Let students work individually or in groups*

Whether individually or in groups, students should be given sufficient time to reason toward an answer. When done individually, students should still be allowed time to share and defend their answers with a neighbor or small group of classmates. This will happen more naturally when students work in small groups, so in both cases, students are given opportunities to articulate their thinking, to become more aware of the thinking of their classmates, to prepare for the classwide discussion, and to develop better listening skills.

One variation is to have students work individually and to submit their answers. The histogram is not displayed yet. Then they work in groups as described above and submit their answers again. This mode permits comparison of the distribution of answers before and after group work, and in some cases can help motivate students who do not appreciate the value of group work. Part of the class-wide discussion can focus on why the distribution of answers changed as a result of working in groups.

The amount of time needed for this phase depends on the item, and sometimes it can be hard to predict how much time a particular class will need for a certain item. Often, the noise level can be a good indicator of where the students are in the process of working on the item. Just after the item is sent, the noise level is often very low, as students are reading and interpreting the item. Then the noise level begins to rise as groups begin to discuss the item. When the noise level has peaked and begins to fall, it is soon time to collect answers.

*Collect answers*

Collecting answers is a good way to get students out of group work smoothly with minimal disruption. It also forces students to commit to an answer—something they will avoid if teachers let them. By committing to an answer, students will be more actively engaged during the other phases of the question cycle, and they will get more out of the experience. When possible, teachers can monitor the answers as they come in, and begin to think about the upcoming class-wide discussion.

*Show histogram*

One of students' favorite parts of the question cycle is the histogram display, in which they get to see how the class is distributed. The immediacy of the feedback helps students see how they fit in to the rest of the class, and serves as an excellent focus for the class-wide discussion. Showing the histogram also returns control of the class to the teacher, and it is done without all the confusion, frustration, and loss of time usually associated with bouncing back and forth between group work and teacher-facilitated activities.

*Discuss as a class*

During this phase, teachers typically ask for volunteers to defend or otherwise explain their choices. Because students have had the opportunity to explain their reasoning in the small-group format, they are often more likely to share their reasoning in the whole-class setting. Because they can see from the histogram that there are classmates who gave the same answer, they are often more comfortable participating, even if they are in the minority. If students are unwilling to share their reasoning, teachers can ask for volunteers who will explain why someone might choose a certain answer, even if it isn't the one they chose.

If the item sent is to predict what will happen during a demonstration, then it is best to do the demonstration <u>after</u> the choices have been discussed. After the demonstration, the discussion should center on any mismatches between predictions and observations.

It is critical to keep the discussion student-directed, even if it is moderated by the teacher. The teacher's role should be to draw out student reasoning and help clarify it. To do this, it is helpful for the teacher to remember that there are no wrong answers and that silence and patience are most likely to instill confidence and willingness to participate. Some paraphrasing can be done, but it is valuable to make sure that the student agrees with the paraphrase and that it is not done so much that students fail to develop appropriate communication skills.

*Wrap-up*

If necessary, and only if necessary, teachers can give a short lecture in order to summarize the discussion, clarify any lingering confusion, introduce useful concepts or principles, emphasize the major points of discussion, indicate connections to earlier questions and discussion, or otherwise address the state of knowledge exhibited by students during the discussion. Another way of wrapping up is to ask students "what if" questions, such as special and limiting cases. An entire question cycle is not needed for these—just a few, short, follow-up questions should suffice. Still another technique is to present the same item again, collect answers, and show a new histogram. The discussion would center on <u>changes</u> in the distribution of answers, and possibly there would be no need of a formal wrap-up.

In some cases, rather than any wrap-up, it is more instructive for students to immediately consider another item, preferably an item that is related to the one just completed. The discussion and wrap-up can then focus on a pair or even a group of items, rather than only on a single item.

**Choosing and sequencing items.**  Before any items can be sent, teachers must select which items they are going to present and decide the order in which they will be presented.  Items should vary in style, difficulty, and purpose.  It is usually best if most of the questions yield a range of responses.  In other words, it is not necessary—and perhaps even undesirable—that questions have only one right answer.  We are looking for questions that will stimulate a lively debate during the class-wide discussion.

Both the habits of mind and the cognitive goals are useful for guiding teachers through this step.  We recommend that teachers first decide which cognitive goal to target within any particular topic or subtopic.  Then, to help maintain interest and motivation, teachers should vary the targeted habit of mind.  However, particular habits of mind are more suitable for certain cognitive goals than others.  For instance, "seeking alternative representations" is good for honing concepts, and "comparing and contrasting" is good for clustering concepts.  Also, a particular habit of mind can advance more than one cognitive goal.

Teachers should remember, however, that they do not need to wait until all the early stages of cognitive development are addressed before moving on to the later stages.  <u>All</u> the cognitive skills should be targeted from the very <u>beginning</u> of the course.  In addition, students do not need to understand every concept, subtopic, or topic completely before moving on to the next.  This means that sometimes students will have to reach and stretch a little.  Teachers should keep in mind that student confusion does not mean that they are lacking information.  A little confusion can be beneficial for providing the internal need that spawns the motivation to learn.

**Preparing for the class-wide discussion.**  Another important step that should be completed before an item is sent is to get ready for the class-wide discussion.  It is common for teachers to be caught off-guard by their students' responses, so some preparation both in the anticipated distribution of answers and in the possible range of explanations will side-step much awkwardness and help maintain a friendly, non-threatening classroom environment.

Teachers must convince students that items are <u>formative</u> not summative.  In other words, the items "inform" rather than "represent a value judgment."  This can be difficult, because students often have years of prior experience trying to determine the "right" answer, and they have lots of emotional "baggage" associated with getting questions right or wrong.  Teachers therefore must disarm students' quest for the right answer, which sometimes happens at the very end of the question cycle, when students want the teacher to summarize the discussion and tell them what the "right" answer was.  This tendency should be avoided, and when students insist on knowing the right answer, teachers should shift the focus of the discussion by *meta-communicating* about why students are so interested in the right answer and why they should be interested instead in the assumptions and thought processes leading up to the answers they chose.

Teachers also must learn to <u>use</u> the information they get back from students.  We expect students to deal with information in real time, so why not teachers as well?  Therefore, teachers need to learn how to listen and respond to students' answers and explanations.

# Collaborative Learning

Although there may be times when teachers would rather have students working individually, we recommend having students work in collaborative groups as much as possible. This does not mean that students get together and hope that someone knows the "right" answer. Collaborative groups are groups in which members are pooling their resources, skills, perspectives, and ideas to reach a common goal, which in this case is not the "right" answer, but increased understanding of every member of the group.

**Skills.** To accomplish this, most students must develop a variety of new interaction and conversational skills, such as listening, interpreting, asking questions, drawing out other points of view, and clarifying. Students must learn how to deal with dominant, passive, uncooperative, and many other personality types. Students must also be willing to participate and to grow in their skills.

Teachers will probably need to spend some time helping students develop the needed skills. Students cannot be expected to learn new material and critical-thinking skills at the same time they are learning new cooperative skills. Special activities must be designed in which the context is familiar and the focus is on interaction and conversation. For instance, possible contexts include basic math, reading, recreational activities, relationships, school, family, and friends. These special activities should be done at the beginning of the year.

Another way to help students develop skills is to give them roles to play, such as Manager (who keeps the team on task), Encourager (who makes sure everyone participates), and Clarifier. Also, some time should be spent having whole-class discussions of how well or poorly group work is going. If students are not willing to participate in this discussion, it may be necessary to create and present an item that focuses on group dynamics. Perhaps after seeing a histogram of results students will be more likely to share their ideas and concerns.

**Advantages of group work.** Teachers might also need to spend some time convincing students that learning these skills is valuable. Even though individual work can be more time efficient (at least in the short run), there are many virtues to having students work in collaborative groups. Students are more likely to become engaged in the learning process when they must explain their point of view to classmates. In other words, when an item is worked on individually, students are more likely to give only minimal thought about the reasons for their answers. Thus, they are ready to submit their answers relatively quickly, but little meaningful learning will occur during the later stages of the question cycle. Group work also gives students an opportunity to hone their arguments and explanations, and gives them increased confidence to share their reasoning during the class-wide discussion.

Finally, group work means that each student's model is being confronted by someone else's. That is, when students disagree, there is hopefully some attempt at resolution, and flaws or weaknesses in a student's model can be uncovered and dealt with. Without collaborative groups, this process requires interaction between the teacher and the student, and it is impossible for the teacher to deal with every individual student. There are fewer manifestations of conflict between the student's model and the teacher's model and there is less learning. With collaborative learning, students can help each other learn, and more students are served.

Developing group skills also has long-term advantages. The former ideal of the idiosyncratic, uncommunicative, competitive individual is becoming less and less desirable. Instead, group skills are increasingly desired for employment, and in some cases, being able to work productively and communicate with others is more highly valued than content knowledge.

**Team size and composition.** A collaborative group can be as small as two students, and in most cases should not be larger than three or four students. The decision about how large the team should be depends mostly on how much time the teacher wants to devote to group work for a given item and on how deeply the teacher wants the students to go with the given item. The more students there are in a team, the wider the range of viewpoints that each student will confront, but the less time each student will have to participate.

The classroom arrangement (i.e., desks, chairs, and tables) can affect the optimal size of a team. If students cannot rearrange themselves so that they are facing each other, then groups should have only 2 or 3 students in them, otherwise one or two students will inevitably become disconnected from the conversation.

Teams can be either *homogeneous* (uniform) or *heterogeneous* (mixed) across a range of categories, such as sex (male/female), achievement (high/medium/low), and math ability. Teams can also be self-selecting, random (i.e., count off or pick numbers from a bowl), or teacher-selected. However, teachers don't really need to worry about team composition for most ASK•IT / A2L items. Two neighbors who simply share their answers and explanations is usually sufficient. This will give each student the maximum amount of engagement and make the transitions into and out of group work as smooth and efficient as possible.

**Other comments on group work.** Even though we encourage teachers to use collaborative groups as much as possible, generally answers are still collected individually. This feature helps to guarantee that students are invested in the outcome and that they can participate fully in the class-wide discussion.

It is useful to watch out for various personality types that might inhibit productive group work. For instance, the *bully* proclaims the "right" answer without explanation and may expect all team members to follow. The bully may even be correct most of the time, but that is wholly irrelevant, because unless the person can explain his/her reasoning behind the answer, the other team members are not helped. Also, bullies need to be convinced that unless they can explain why an answer is correct, they do not really understand it.

*Holdouts* refuse to tell anyone their answers, often because they do not want to share their

16                                             Ask•IT / A2L: Assessing Student Knowledge with Instructional Technology

expertise with anyone else and thereby help them. These students can be too competitive and may be trying too hard to get the best grade—at the expense of everyone else. Some even consider it cheating to share their answers and explanations with others. All of these holdouts need to be convinced that sharing ideas is a critical factor in maximizing everyone's learning.

Meta-communication is a particularly useful mode for dealing with strains in the collaborative process. When students become disengaged or disenfranchised, it can be used to get everyone involved again. When there are personality conflicts and when groups are not functioning optimally, it can be used to get the groups back on track again. It can be used to discuss how to work in groups, to permit students to express their opinions about group work, and to explain the value of group work for learning and developing useful skills.

## The Role of Technology

The ASK•IT / A2L approach can be supported and enhanced using technology (in the form of a classroom communication system) to monitor individual or group progress on an item, collect students' answers, compile and sort the answers into histograms, display the histograms, and store responses for possible retrieval at a later time. Without technology, it would be difficult—but not impossible—to duplicate many of the key ingredients that lead to student involvement and teacher awareness.

**Monitoring progress on a given item.** One type of item is a group of questions or tasks that students work on in sequence, submitting answers to each before moving on to the next. Some technology makes it possible to monitor which groups are going ahead of and which are lagging behind the rest. Knowing which groups are lagging behind permits the teacher to target the groups that need attention and to spend extra time working with them. Thus, technology can make group work more efficient by avoiding the situation in which everyone is waiting for one or two groups to finish. Technology used in this way also avoids the situation in which some groups are not allowed to finish because there is no more time. This option can be devastating, because often the groups lagging behind are those that are struggling the most with the material, and if they don't finish, they will not be able to fully participate in the class-wide discussion.

**Collecting, compiling, and sorting answers.** Technology can make the collection of answers extremely efficient and easy. With technology, this phase usually takes about a minute and requires relatively little effort by both students and teachers. Students just submit their answers using some sort of input device, and the phase is over. Without technology, students must use worksheets, bubble sheets, or flash cards to submit their answers.

Using flash cards is the easiest and most efficient of these other options and therefore would seem to be the preferred choice, but it is not anonymous, which many students do not like. Also, using flash cards promotes the teacher-centered approach, because only the teacher can see all the cards, and therefore, only the teacher is getting the complete picture. We want all students to be aware of how their classmates answer every question, so if flash cards are used, a step should be added. For instance, a student could tally the results and write them on the blackboard for the rest of the class to see.

Compiling and sorting answers is also tedious and time-consuming without technology. Worksheets, answer sheets, and bubble sheets need to be processed, so the results might not be available immediately or it might take valuable class time to get the results needed to display the histogram.

**Displaying the histogram.** This phase of the question cycle is one of the most important, because it serves so many purposes. It lets the teacher know how the <u>whole</u> class is thinking about a given item. It provides immediate feedback to students and shows them how they fit in to the class. It serves as an ideal intermediate stage between group work and the class-wide discussion. It also serves as a good starting point for the class-wide discussion. Even if a student is in the minority, he or she is often more likely to participate in the whole class discussion because they know that some of their classmates chose the same answer.

**Storing responses.** The technology also keeps track of how each student is doing by storing their answers to each item. This means that teachers can retrieve this information at any time and see how individual students are progressing. If necessary, teachers can then take steps to help students. Teachers can also use the information to make long-term decisions about what types of items to select in the future.

**Added value.** There are a number of additional benefits of using technology. It is an excellent time-management tool and helps the teacher gently move students from one phase of the lesson to the next. For instance, typically there is a time limit given for submission of the answers. This time limit is set by the instructor but enforced by the technology. During the time period when the system is accepting answers, students must bring their attention back to the classroom so that they can get their answers submitted in time.

Also, most students enjoy using technology; it's new; it's fun and exciting. Even students who are intimidated by computers are not generally intimidated by this use of technology. Finally, classroom technology is becoming more common at both the high school and college levels and in many different disciplines. Some experience with technology will help students get up to speed in these other courses.

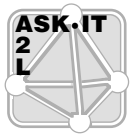

# PART IV.
# Developing New Items

Teachers using the ASK•IT / A2L approach will inevitably need to create their own questions, problems, and tasks. In order to help teachers develop new items in keeping with the desired approach and outcomes, we offer the following model-based design paradigm.

## Model-Based Design Paradigm

To develop items, we use the *model-based design paradigm* shown graphically below. We start with findings from many different strands of cognitive research as well as three premises about the role of cognition in developing problem-solving proficiency. We use these to develop a framework for the acquisition and storage of knowledge and the development of analysis, reasoning, and problem-solving skills. From the cognitive framework, we derive our instructional approach—concept-based problem solving. And finally, we use a "web" or "matrix" of cognitive goals and habits of mind to create instructional materials for all of our projects— Minds•On Physics, ASK•IT, and Assessing-to-Learn Physics. In other words, for each cognitive goal, we create items to encourage one or more habits of mind. The result is a rich set of items that shift the focus of instruction from pure problem-solving tasks to activities that promote beneficial mental processes.

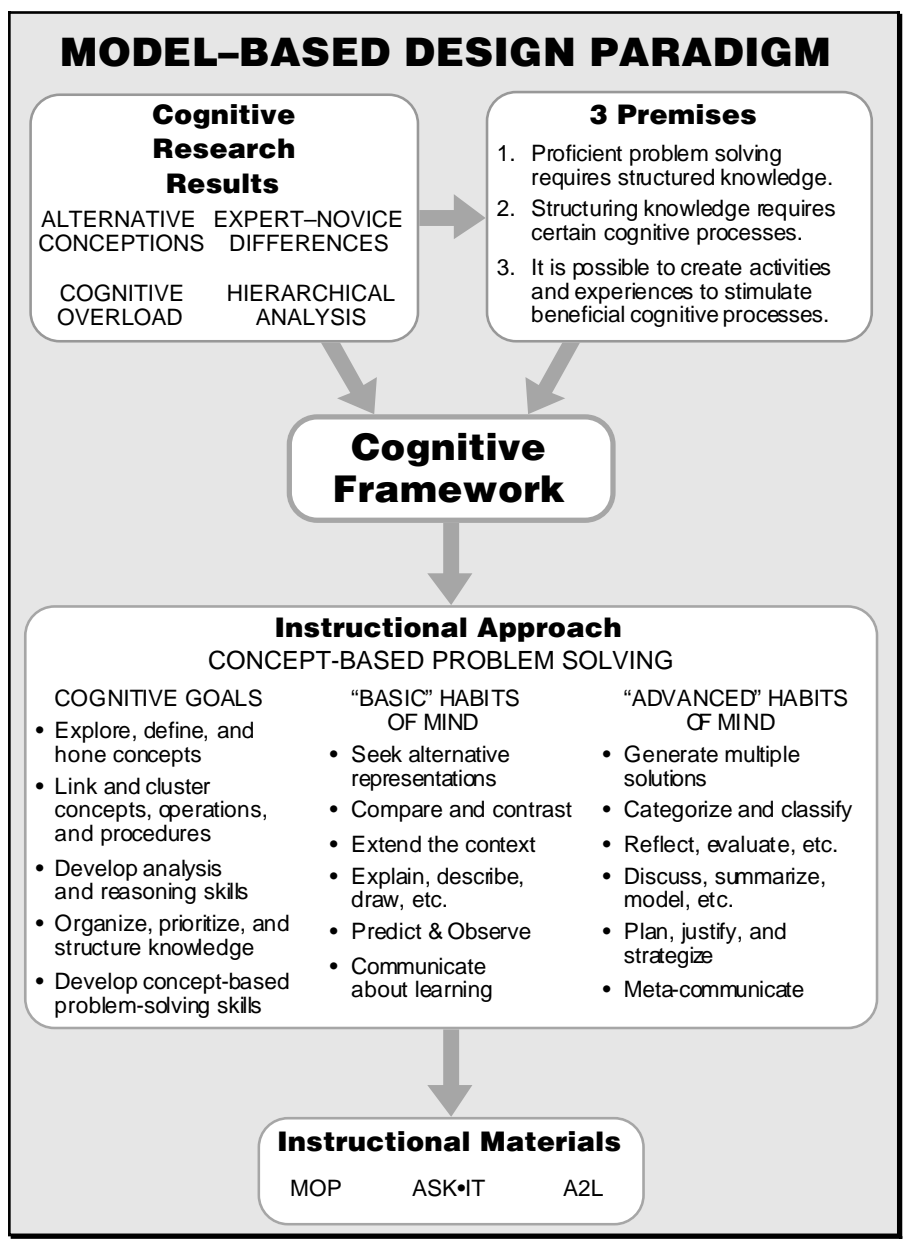

# Applying the Model-Based Design Paradigm
# to Create New ASK•IT / A2L Items

There is no cookbook for developing items. The most common pitfall occurs when teachers merely copy the superficial features of a successful item without really understanding why the item was successful. In the end, practice, testing, observation, and reflection will be the most useful guide to developing new items.

**Modeling students.** Having a model for how students answer questions will help. Teachers should listen to students' answers and explanations, and develop the skill of paraphrasing them, always checking with students to make sure they have paraphrased accurately. Teachers can predict the outcome of an item and try to guess what some of the explanations will be. They can think about the factors that affect the accuracy of the predictions. All of this will help teachers model their students.

**Healthy attitudes.** Having a try-anything attitude will help also. It might take many attempts—and numerous failures—before teachers consistently create successful items. Time and practice will lead to improvement, especially if teachers are reflecting on their development and use of items. Unfortunately but unavoidably, teachers must practice in front of a large, at times unsympathetic, audience: their students! So, teachers who are less concerned about "getting it right" and more concerned with learning will improve the quickest. (But isn't this attitude we are trying to instill in our students as well?)

Teachers also should exhibit the behaviors and habits of mind they value and expect from their students: explore new contexts, predict and observe, seek alternative representations, compare and contrast, meta-communicate (with other teachers), and so on. They can use the feedback they get from items to hone their understanding and appreciation of formative assessment.

**Using the model-based design paradigm.** The model-based design paradigm should help teachers organize their thoughts, observations, predictions, and new creations. For each item, it is useful to know which habit(s) of mind and which cognitive goal(s) it is targeting. (Actually, most items target more than one habit of mind and more than one cognitive goal.)

In order to develop new items, we often use the habits of mind as instructional modes. For instance, to encourage students to consider alternative representations such as graphs, we ask numerous questions asking them to interpret, draw, and translate graphs. These items serve to explore, define, hone, and link concepts. They also link concepts to certain operations and procedures, such as interpreting the slope of a graph and the area beneath a graph. We ask questions that are most easily answered using graphs, thereby developing analysis, reasoning, and problem-solving skills. We then ask students to discuss and reflect on the item, perhaps to find out why they did not consider a graph to answer the question, and then meta-communicate with them to convince them of the value of graphs for analysis, reasoning, and problem solving.

Some habits of mind are more appropriate for certain cognitive goals than others. The following table shows which habits of mind are most appropriate for each cognitive goal.

| Habit of Mind | Explore, define, and hone concepts | Link and cluster concepts, etc. | Develop analysis and reasoning skills | Develop problem-solving skills | Organize, prioritize, and structure knowledge |
|---|---|---|---|---|---|
| Seek alternative representations | ✧ | ✧ | ✧ | ✧ | ✧ |
| Extend the context | ✧ | | ✧ | ✧ | |
| Compare and contrast | ✧ | ✧ | ✧ | ✧ | ✧ |
| Explain, draw, describe, etc. | ✧ | ✧ | ✧ | ✧ | ✧ |
| Predict & Observe | ✧ | ✧ | ✧ | | |
| Monitor / refine communication | ✧ | ✧ | | | ✧ |
| Generate multiple solutions | | | ✧ | ✧ | ✧ |
| Reflect, evaluate, etc. | ✧ | ✧ | ✧ | ✧ | ✧ |
| Categorize and classify | | ✧ | ✧ | ✧ | ✧ |
| Discuss, model, summarize, etc. | ✧ | ✧ | ✧ | ✧ | ✧ |
| Plan, justify, and strategize | | | ✧ | ✧ | ✧ |
| Meta-communicate | ✧ | ✧ | ✧ | ✧ | ✧ |

*Exploring, defining, and honing concepts*

The basic habits of mind are particularly good for developing concepts and representations of concepts. Use a familiar situation, or use the same situation to ask a few questions. Compare situations. Encourage students to use concepts to explain their predictions and describe their observations.

The key is to use questions—rather than lecturing—to get students to realize that their definitions might be weak, imprecise, inconsistent, and/or incompatible with those of their classmates. Questions should create a need to know and the motivation to adopt the accepted definitions. For instance, give students five situations, and ask them which ones (or how many) have "potential energy" in them. Or ask them if the "energy" of liquid water put into a freezer is increasing, decreasing, or staying the same. They don't need to have learned the definitions or have heard the expressions before. They will tell you their definitions. When there is conflict and a need to know, they become ready to learn the physics definition.

*Linking and clustering concepts, operations, and procedures*

To link concepts to each other, students should appreciate the relationships between concepts. For instance, they must appreciate that the slope of velocity vs. time is acceleration vs. time. Therefore, give students a graph of velocity vs. time and ask questions about the acceleration vs. time: When is it maximum? When is it minimum? When is it largest? When is it smallest? When is the velocity changing the most? When is the velocity changing the least? These questions will not only link the concepts of velocity and acceleration, but will also help students make distinctions between speed, velocity, acceleration, and average acceleration.

To link concepts to operations and procedures, students should recognize the role that concepts play in deciding the applicability and usefulness of operations and procedures to analyze, reason, and problem solve. For instance, finding the components of vectors is needed to find the net force, which is needed to apply Newton's 2nd law. Therefore, ask questions involving concepts that require the use of common operations and procedures. For instance, have students compare the normal forces exerted on two crates being pulled along a floor by strings having the same tension but held at different angles.

To cluster concepts, operations, and procedures, students should classify and categorize what they have learned. Therefore, give students a set of four or five concepts, operations, and procedures, and ask students which one does not go with the rest. Have students defend their choices. (Most likely, every choice will be defensible! And we can predict this outcome without knowing what the set contains, because every student will perceive a unique pattern in the given set!!) Have students debate their choices. If desired, teachers can wrap-up by saying that physicists generally categorize according to what is more widely applicable or useful for reasoning, analysis, and problem solving. For instance, we typically use "Newton's laws" to organize one set of concepts, operations, and procedures, and "Conservation of Energy" to organize another.

*Developing analysis and reasoning skills*

Almost all of the habits of mind are useful instructional modes for developing analysis and reasoning skills, primarily because the habits of mind have such a close association with these skills. The habits of mind are the core tendencies we are trying to encourage, and analysis and reasoning are the core skills we are trying to develop. If students are exhibiting one of these behaviors, either voluntarily as a habit of mind or during an ASK•IT / A2L item as an instructional mode, then they are improving their ability to analyze and reason about physical situations.

*Developing concept-based problem-solving skills*

The "advanced" habits of mind are particularly good for developing concept-based problem-solving skills. Have students indicate how many different ways they can solve a particular problem, then have them defend their choices. Give students an invalid plan for solving a problem, and ask them to indicate which step in the plan is invalid. Ask students which principles can be applied to a given situation. Have the class list the top five or 10 most difficult aspects of solving problems, then have each individual indicate which one is hardest for him/her.

*Organizing, prioritizing, and structuring knowledge*

The "advanced" habits of mind are also particularly good instructional modes for structuring knowledge. Ask students to sort ideas. Ask them to consider different concept maps and indicate which is most like the way they think about physics. Have students evaluate different organizational schemes and indicate which they like best and worst.

**Final thoughts.** Teachers should keep in mind that to help students develop concepts, skills, and representations, the situation should be familiar, or if unfamiliar, they should use the same situation many times. To check to see if concepts, skills, or representations have been learned, the situation or representation should be unfamiliar, or if familiar, it should be different from the situations and representations used to learn the concepts, skills, and representations. Many teachers tend to do the opposite: they use unfamiliar situations while students are learning—a cognitively demanding task—then use the same situations to check for understanding—which leads to superficial learning.

Teachers should not be afraid of using the same situation over and over again. This will reduce cognitive overload for students. Students will become familiar with the situation, and as a result, the class will be able to achieve greater subtlety. It will encourage higher level cognitive skills and deeper cognitive growth.

Another teacher tendency is to ask questions with only one defensible answer, which tests and encourages only low level skills. Teachers should try to develop items that have more than one defensible answer—answers that depend on the assumptions students make. The broader the distribution of answers, the better.

Teachers also often give items in which all and only the needed information is given. Students can become drowned in information and quickly learn that the answer will be a simple manipulation of the given information. Therefore, wherever possible, minimize the amount of information given in an item, and encourage students to think about how they will determine any needed information. Also, sometimes give students extraneous information, so that they learn to organize and prioritize information, and discard what they don't need.

It may take a long time for teachers to acquire the skill of developing new items. Therefore, they should try to have fun creating and trying out questions. Experience, and reflecting on experience, are often the best guides to success.

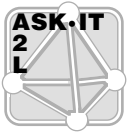

# PART V.
# Metacognition

The habits of mind described in part II are not limited to students. Teachers need to improve theirs as well. The difference is that the habits of mind are used to analyze and solve the problem of instruction. But we do not need to stop here either. The habits of mind are also applicable to students learning about learning, and they are applicable to teachers learning about teaching.

## Applying Habits of Mind to Teachers

**Basic habits of mind.** The table below shows examples of how teachers can apply the six basic habits of mind to their own instructional practice. Note that the examples apply not only to teachers interactions with students but also to their interactions with colleagues.

| Habit of Mind | Teachers can... |
|---|---|
| Seek alternative representations | • seek alternative explanations and interpretations of students' responses.<br>• consider different explanations for what motivates students.<br>• talk to other teachers about their models for student behavior.<br>• challenge assumptions. |
| Extend the context | • be more experimental with classroom techniques.<br>• be willing to change the classroom dynamic when it is not working.<br>• shift the focus of instruction from covering content to monitoring process skills. |
| Compare and contrast | • contrast their new or evolving approach with their old one.<br>• compare their approach with that of another teacher.<br>• be willing to be observed by another instructor. |
| Explain, describe, draw, etc. | • be willing to interact with individual students.<br>• develop better conversational skills.<br>• explain their instructional goals to students.<br>• defend their instructional choices, such as collaborative learning.<br>• describe to a colleague something that happened during class. |
| Predict & Observe | • predict the result of an interaction (based on their models of students).<br>• observe the result of an interaction.<br>• predict what will happen by being experimental.<br>• observe the result of changing the classroom dynamic.<br>• observe a colleague. |
| Monitor and refine communication | • pay attention to how well or poorly students are interacting with each other.<br>• resist jumping to conclusions.<br>• listen.<br>• ask questions.<br>• consider that students may be using terms differently than expected.<br>• try to get colleagues to talk about teaching.<br>• develop a vocabulary to discuss educational issues with colleagues and friends. |

**Advanced habits of mind.** The following are examples of how teachers can apply the six advanced habits of mind to their teaching:

| Habit of Mind | Teachers can... |
|---|---|
| Generate multiple solutions | <ul><li>consider different approaches to instruction.</li><li>try different classroom dynamics/communication modes.</li><li>apply different analogies to teaching, such as parenting, coaching, etc.</li><li>generate more than one lesson plan.</li></ul> |
| Reflect, evaluate, etc. | <ul><li>look for patterns between their teaching styles and student behavior and involvement.</li><li>look for patterns between a colleague's teaching style and student outcomes.</li><li>take time to reflect daily on practice and communication.</li><li>evaluate their communication/conversational skills.</li></ul> |
| Categorize and classify | <ul><li>organize their behaviors into those that foster student involvement and those that do not.</li><li>classify students according to their learning styles.</li><li><u>resist</u> categorizing students according to superficial features, such as sex, racial group, attention span, etc.</li><li>use habits of mind to classify students.</li><li>use stages of cognitive development to structure the school year.</li><li>develop categories to explain student behavior.</li></ul> |
| Discuss, summarize, model, etc. | <ul><li>talk to students about their instructional approaches.</li><li>regularly summarize their instructional goals for their students.</li><li>discuss their experiences with colleagues.</li><li>use analogies, such as coaching and parenting, to explain the reasons behind their classroom practices.</li><li>form models of students; discuss their models with others.</li><li>listen to other teachers' models of students.</li><li>organize an "action research" group with local physics teachers.</li></ul> |
| Plan, justify, and strategize | <ul><li>justify their lesson plans with, e.g., parenting principles.</li><li>form intermediate goals for students and develop a plan to achieve them.</li><li>develop a strategy for dealing with communication and motivation issues.</li><li>create a plan for developing a communication-based instructional style.</li><li>discuss their strategies for developing new skills with other teachers.</li></ul> |
| Meta-communicate | <ul><li>talk to students about the skills needed to thrive, learn, and succeed.</li><li>talk to students about the value and importance of structured knowledge.</li><li>talk to the class about the value of collaborative group work.</li><li>when the class is being uncooperative or uncommunicative, or when students are being disruptive or are not participating, stop the lesson and discuss it.</li><li>talk to other teachers about habits of mind and stages of cognitive development.</li></ul> |

The key to both of these lists of examples is that teachers benefit from having and/or acquiring the same habits of mind that students benefit from having. The difference is that a teacher applies the habits of mind to the problem of communication and learning.

# Applying Habits of Mind to Students' Metacognition

Students also benefit from applying the habits of mind to their own learning, communication skills, and cognition. In other words, the habits of mind apply equally well to the "problem" of students learning how to learn, think, and communicate.

**Basic habits of mind.** The table below shows examples of how students can apply the basic habits of mind to their learning.

| Habit of Mind | Students can… |
|---|---|
| Seek alternative representations | - consider other points of view.<br>- determine if they are using the same definitions as others.<br>- look up unknown terms in a dictionary.<br>- view the teacher as being on the same side as they are, i.e., rather than being in conflict.<br>- see interactions with classmates as helpful and cooperative, rather than distracting, competitive, or a waste of time.<br>- translate the teacher–student relationship into a parenting or coaching model. |
| Extend the context | - participate during the class-wide discussion.<br>- try to apply principles learned in class to everyday situations.<br>- apply thinking skills learned in physics to other courses.<br>- apply communication skills to other areas, e.g., relationships with friends and family. |
| Compare and contrast | - compare their learning styles to that of their classmates.<br>- contrast their levels of participation to that of their classmates.<br>- compare their teacher's instructional approach with that of other teachers.<br>- keep track of any changes in their learning styles and approaches.<br>- keep track of any changes in their reasoning, analysis, or communication skills. |
| Explain, describe, draw, etc. | - describe their learning styles to their classmates or teacher.<br>- explain their studying habits.<br>- describe the role of communication in learning.<br>- assess the extent to which they are interesting in learning, as opposed to, e.g., getting a good grade or passing an AP test.<br>- identify their long-term goals.<br>- describe their views on education. |
| Predict & Observe | - predict the distribution of students' answers to a question.<br>- predict the reasons for other students' answers.<br>- defend their predictions concerning other students' answers.<br>- observe which classmates are more/less willing to participate and engage.<br>- observe which skills the teacher is trying to encourage. |
| Monitor and refine communication | - make sure the teacher is talking about communication skills.<br>- talk about group dynamics—successes, failures, dominant personalities, etc.<br>- seek advice about ways to improve communication.<br>- see if their classmates are paying attention to each other during the class-wide discussions. |

**Advanced habits of mind.** The following are examples of how students can apply the advanced habits of mind to their learning, thinking, and communication.

| Habit of Mind | Students can... |
|---|---|
| Generate multiple solutions | - attempt more than one method for learning something.<br>- try different modes of communication.<br>- develop different roles for collaborative learning. |
| Reflect, evaluate, etc. | - look for patterns in how their classmates participate, learn, and communicate.<br>- evaluate their own communication skills.<br>- assess their own thinking skills.<br>- think about learning, thinking, and communication.<br>- reflect on how well or poorly they communicated during a particular activity, such as a class-wide discussion or collaborative-group activity. |
| Categorize and classify | - categorize their classmates according to their learning styles.<br>- classify their learning experiences according to how well or poorly they learned.<br>- organize modes of interaction (lecture, collaborative group work, in-class discussion, etc.) according to how well or poorly communication is achieved or maintained.<br>- analyze behaviors using the habits of mind as organizing concepts. |
| Discuss, summarize, model, etc. | - discuss with their friends the role of communication in relationships.<br>- form models of how classmates think about learning.<br>- form models of how teachers think about instruction.<br>- paraphrase their classmates' ideas about learning and thinking.<br>- listen to a conversation about the most important ideas or skills needed to do physics.<br>- summarize a discussion about the habits of mind needed for success in life. |
| Plan, justify, and strategize | - go into a lesson with an agenda—a set of desired outcomes and a plan for achieving them.<br>- take a more strategic approach to learning, e.g., decide which skills need to be honed and develop a plan for doing it.<br>- justify the habits of mind.<br>- create a plan for improving communication or promoting interaction.<br>- defend their learning styles. |
| Meta-communicate | - talk to classmates about why it is important and useful to discuss learning, thinking, and communication.<br>- discuss the features of good communication and interaction.<br>- develop a vocabulary for discussing thinking and learning.<br>- talk about the difficulties in and hindrances to discussing communication and its impact on learning and thinking.<br>- discuss ways to motivate classmates to talk about learning and thinking. |

Students are not inclined to apply any of these habits of mind to learning physics, so they will be even less inclined to apply them to learning about learning, thinking, and communication. However, teachers can overcome some of students' reluctance by incorporating these examples into their lessons and by convincing them that there is nothing as vital, important, useful, or interesting as one's own thought processes.

# CONCLUSION

There are many ironies in science instruction. To learn concepts, for example, it is not sufficient to drill and practice learning definitions and computing values. Understanding concepts requires using them in a wide variety of contexts. To learn problem solving, it is not sufficient to do lots of problems. To be able to solve problems flexibly, a student must understand concepts, principles, and procedures, and must be able to reason and analyze as well. They must reflect on the problem-solving process, and find patterns that are useful to them for organizing and prioritizing knowledge.

Learning how to play a sport can be a useful analogy for learning science. For instance, running is analogous to thinking. Young players often have limited stamina and may have a tendency to resist running, just as students resist thinking. But a coach would insist on the players running to build up their stamina, and the players would eventually recognize the value of running and that running gets easier with practice. Teachers, too, must insist that their students reason and analyze, because learning—true learning—is impossible without the ability to reason and analyze.

Parenting is also a useful analogy. Teachers are not going to be there to interpret every question, provide all needed information, or make every decision in a student's life. Therefore, students must develop the skills to think independently and critically about the issues that touch their lives.

The ASK•IT / A2L project seeks to help students and teachers develop habits of mind that will help them to cope with problems, questions, and situations they face in their daily lives, both in school and beyond school. By asking students to consider open-ended, multi-faceted tasks and by avoiding certain common classroom practices, teachers can instill in students a need to know and the motivation to learn. By focusing on moderate goals, rather than short-term goals, teachers can achieve the long-term goals they usually desire, such as deep conceptual understanding, analysis and reasoning skills, and flexible problem-solving skills.

Some of you might agree in principle to the virtues and desirable aspects of the ASK•IT / A2L approach but think that the goals are unrealistic or impractical. We have found that students respond intuitively to their teachers' attitudes about them. Teachers' views of students and instruction can impact—either positively or negatively—students' interest and motivation. This means that we should not give up on the approach and its ideals too easily, because then we could be giving up on our students and their futures.